\documentclass[12pt]{article}

\usepackage{amsmath,amssymb}

\global\arraycolsep=1pt
\numberwithin{equation}{section}
\newcommand{\bel}[1]{\begin{equation}\label{#1}}                     
\newcommand{\bal}[1]{\begin{eqnarray}\label{#1}}                     
\newcommand{\be}{\begin{equation}}
\newcommand{\ee}{\end{equation}}

\newcommand{\rmd}{\mathrm{d}}
\newcommand{\rmi}{\mathrm{i}}

\newcommand{\scr}{\scriptstyle}
\newcommand{\qq}{\qquad}

\renewcommand{\thefootnote}{\fnsymbol{footnote}}

\begin{document}

%%%%%%%%%%%%%%%%%%%%%%%%%%%%%%%%%%%%%%%%%%%%%%%%%%%%%%%%%%%%%%%%%%%%%%%%%%%%%%%%%%%%%%%%%%%%%%%%%%

\begin{flushright}
OCU-PHYS 258 \\
hep-th/0611285
\end{flushright}
\vspace{20mm}

\begin{center}
{\bf\Large
Kerr-NUT-de Sitter Curvature\\
in All Dimensions
}
\end{center}

\begin{center}

\vspace{15mm}

Naoki Hamamoto$^a$\footnote{
\texttt{hamamoto@sci.osaka-cu.ac.jp}
},
Tsuyoshi Houri$^a$\footnote{
\texttt{houri@sci.osaka-cu.ac.jp}
}, 
Takeshi Oota$^b$\footnote{
\texttt{toota@sci.osaka-cu.ac.jp}
} and
Yukinori Yasui$^a$\footnote{
\texttt{yasui@sci.osaka-cu.ac.jp}
}

\vspace{10mm}

${}^a$
\textit{
Department of Mathematics and Physics, Graduate School of Science,\\
Osaka City University\\
3-3-138 Sugimoto, Sumiyoshi,
Osaka 558-8585, JAPAN
}
\vspace{5mm}

\textit{
${{}^b}$
Osaka City University
Advanced Mathematical Institute (OCAMI)\\
3-3-138 Sugimoto, Sumiyoshi,
Osaka 558-8585, JAPAN
}

\vspace{5mm}

\end{center}
\vspace{8mm}

%%%%%%%%%%%%%%%%%%%%%%%%%%%%%%%%%%%%%%%%%%%%%%%%%%%%%%%%%%%%%%%%%%%%%%%%%%%%%%%%%%%%
\begin{abstract}
We explicitly calculate the Riemannian curvature of $D$-dimensional metrics recently
discussed by Chen, L\"{u} and Pope.
We find that they can be concisely written by using a single function.
The Einstein condition which corresponds to the Kerr-NUT-de Sitter metric
is clarified for all dimensions.
It is shown that the metrics are of type D.

\end{abstract}
%%%%%%%%%%%%%%%%%%%%%%%%%%%%%%%%%%%%%%%%%%%%%%%%%%%%%%%%%%%%%%%%%%%%%%%%%%%%%%%%%%%%%

\vspace{25mm}

\newpage

\renewcommand{\thefootnote}{\arabic{footnote}}
\setcounter{footnote}{0}

%%%%%%%%%%%%%%%%%%%%%%%%%%%%%%%%%%%%%%%%%%%%%%%%%%%%%%%%%%%%%%%%%%%%%%%%%%%%%%%%%%%%%%%%%%%%%%%%%%%%%%%%%%
\section{Introduction}

In recent years, studies of exact solutions to the higher dimensional
Einstein equations have attracted much attention in the
context of supergravity and superstring theories \cite{MP,HHT,GLPP1,GLPP2,CLP1,CLP2}.
Here, we revisit a class of $D$-dimensional metrics discussed by Chen, L\"{u} and Pope \cite{CLP2}~:\\
(a) $D=2 n$ 
\begin{equation}
g = \sum_{\mu=1}^{n} \frac{\rmd x_{\mu}^2}{Q_\mu}+
\sum_{\mu=1}^{n} Q_{\mu} \left( \sum_{k=0}^{n-1} A_{\mu}^{(k)}
\rmd \psi_k \right)^2~~, 
\end{equation}
(b) $D=2 n+1$
\begin{equation}
g = \sum_{\mu=1}^{n} \frac{\rmd x_{\mu}^2}{Q_\mu}+
\sum_{\mu=1}^{n} Q_{\mu} \left( \sum_{k=0}^{n-1} A_{\mu}^{(k)}
\rmd \psi_k \right)^2+S\left( \sum_{k=0}^{n} A^{(k)}
\rmd \psi_k \right)^2~~. 
\end{equation}
The functions $Q_{\mu}~(\mu=1,2, \cdots,n)$ are given by
\begin{equation}
Q_\mu= \frac{X_\mu}{U_\mu},\qquad \qquad
U_{\mu} = \prod_{\stackrel{\scr \nu=1}{(\nu \neq \mu)}}^n
( x_{\mu}^2 - x_{\nu}^2 ),
\end{equation} 
where $X_\mu$ is an arbitrary function
depending only
on $x_\mu$. The remaining functions are
\begin{equation}
A_{\mu}^{(k)}
= \sum_{\stackrel{\scr 1 \leq \nu_1 < \nu_2 < \dotsm
< \nu_k \leq n}{(\nu_i \neq \mu)}}
x_{\nu_1}^2 x_{\nu_2}^2 \dotsm x_{\nu_k}^2,~~
A^{(k)}
= \sum_{1 \leq \nu_1 < \nu_2 < \dotsm
< \nu_k \leq n}
x_{\nu_1}^2 x_{\nu_2}^2 \dotsm x_{\nu_k}^2,
\end{equation}
($A_\mu^{(0)}=A^{(0)}=1$) and $S=c/A^{(n)}$ with a constant $c$.
It has been shown by means of
computer calculation that the metrics satisfy the Einstein
equations $Ric(g)= \Lambda g$
for dimensions $D \le 15$ if $X_\mu$ takes the form~:\\
(a) $D=2 n$
\begin{equation}
X_{\mu}=\sum_{k=0}^{n} c_{2 k} x_{\mu}^{2 k}+b_{\mu} x_{\mu}, 
\end{equation}
(b) $D=2 n+1$
\begin{equation}
X_{\mu} =\sum_{k=1}^{n} c_{2 k} x_{\mu}^{2 k}+b_{\mu}+
\frac{(-1)^{n}c}{x_{\mu}^2}, 
\end{equation}
where $c, c_{2 k}$ and $b_{\mu}$ are
free parameters. 
This class of metrics gives the Kerr-NUT-de Sitter
metric \cite{CLP2}, and the solutions in \cite{MP,HHT,GLPP1,GLPP2,CLP1} 
are recovered by choosing special parameters.
However, the explicit Riemannian curvature  was not given in their analysis. So
it is obscure how the metrics become solutions to the Einstein equations. In this paper
we give a systematic investigation of the Riemannian curvature.
We show that they can be concisely written by using a single function.
We also prove that the metrics (1.1) and (1.2) are  of type D and 
under the conditions (1.5) and (1.6) they becomes Einstein metrics for all dimensions.
This family of metrics are also
interesting
from the point of view of AdS/CFT correspondence. 
Indeed, odd dimensional Einstein metrics lead to
Sasaki-Einstein metrics by taking the BPS limit \cite{HSY,CLPP1,CLPP2,CLP2} 
and even dimensional Einstein metrics lead to
Calabi-Yau metrics in the limit \cite{CLP2,OY,LP}. Especially,
the five dimensional Sasaki-Einstein
metrics have emerged quite naturally in the 
AdS/CFT correspondence.

%%%%%%%%%%%%%%%%%%%%%%%%%%%%%%%%%%%%%%%%%%%%%%%%%%%%%%%%%%%%%%%%%%%%%%%%%%%%%%%%%%%%%%%%%%%%%%%%%%%%%%%%%%%%%%%%%%%
\section{$D=2n$}

For the metric (1.1) we introduce the following orthonormal frame 
$\{ e^a \}=\{ e^{\mu}, e^{n+\mu} \}$ $(\mu=1,2,\dotsc, n)$:
\begin{equation}
e^{\mu} = \frac{\rmd x_{\mu}}{\sqrt{Q_{\mu}}}, \quad
e^{n+\mu} = \sqrt{Q_{\mu}}
\left( \sum_{k=0}^{n-1} A_{\mu}^{(k)} \rmd \psi_k \right).
\end{equation}
Using the first structure equation
\begin{equation}
\rmd e^a + \omega^a{}_b \wedge e^b = 0
\end{equation}
and $\omega_{ab} = - \omega_{ba}$, we obtain 
connection $1$-forms $\omega_{ab}$. A straightforward calculation
gives
\be\begin{split}
\omega_{\mu \nu} &=
( 1 - \delta_{\mu \nu} ) 
\left[ 
- \frac{x_{\nu} \sqrt{Q_{\nu}}}{x_{\mu}^2 - x_{\nu}^2} e^{\mu} 
- \frac{x_{\mu} \sqrt{Q_{\mu}}}{x_{\mu}^2 - x_{\nu}^2} e^{\nu} 
\right], \cr
\omega_{\mu, n+\nu} &=
\delta_{\mu \nu}
\left[ - \frac{\partial( \sqrt{Q_{\mu}} )}{\partial x_{\mu}}
e^{n+\mu} + \sum_{\stackrel{\scr \rho=1}{(\rho \neq \mu)}}^n
\frac{x_{\mu} \sqrt{Q_{\rho}}}{x_{\mu}^2 - x_{\rho}^2}
e^{n+\rho} \right] \cr
& + ( 1 - \delta_{\mu \nu})
\left[ - \frac{x_{\mu} \sqrt{Q_{\mu}}}{x_{\mu}^2 - x_{\nu}^2}
e^{n+\nu} 
+ \frac{x_{\mu} \sqrt{Q_{\nu}}}{x_{\mu}^2 - x_{\nu}^2}
e^{n+\mu} \right], \cr
\omega_{n+\mu, n+\nu}
&= (1 - \delta_{\mu \nu})
\left[ - \frac{x_{\mu} \sqrt{Q_{\nu}}}{x_{\mu}^2 - x_{\nu}^2}
e^{\mu}
- \frac{x_{\nu} \sqrt{Q_{\mu}}}{x_{\mu}^2 - x_{\nu}^2}
e^{\nu} \right].
\end{split}
\ee
From the second structure equation
\be
R_{ab} = \rmd \omega_{ab} + \omega_{ac} \wedge \omega^c{}_b,
\ee
we can calculate the curvature $2$-forms $R_{ab}$. 
It is convenient to introduce a quantity
\be
Q_T= \sum_{\mu=1}^n Q_{\mu}.
\ee
We find $(\mu \neq \nu)$
\be
\begin{split}
R_{\mu\nu}
&= - \frac{1}{2(x_{\mu}^2 - x_{\nu}^2)}
\left( x_{\mu} \frac{\partial Q_T}{\partial x_{\mu}}
- x_{\nu} \frac{\partial Q_T}{\partial x_{\nu}} \right)
e^{\mu} \wedge e^{\nu} \cr
& \qq - \frac{1}{2(x_{\mu}^2 - x_{\nu}^2)}
\left( x_{\nu} \frac{\partial Q_T}{\partial x_{\mu}}
- x_{\mu} \frac{\partial Q_T}{\partial x_{\nu}}
\right) e^{n+\mu} \wedge  e^{n+\nu},\cr
R_{\mu, n+\mu} &= 
- \frac{1}{2} \frac{\partial^2 Q_T}{\partial x_{\mu}^2}
e^{\mu} \wedge e^{n+\mu} \cr
& \qq + \sum_{\rho \neq \mu}
\frac{1}{x_{\mu}^2 - x_{\rho}^2}
\left( x_{\mu} \frac{\partial Q_T}{\partial x_{\rho}}
- x_{\rho} \frac{\partial Q_T}{\partial x_{\mu}} \right)
e^{\rho} \wedge e^{n+\rho},
\qq (\mbox{no sum})\cr
R_{\mu, n+\nu} &= 
- \frac{1}{2(x_{\mu}^2-x_{\nu}^2)}
\left( x_{\mu} \frac{\partial Q_T}{\partial x_{\mu}}
- x_{\nu} \frac{\partial Q_T}{\partial x_{\nu}} \right)
e^{\mu} \wedge e^{n+\nu} \cr
& + \frac{1}{2(x_{\mu}^2 -x_{\nu}^2)}
\left( x_{\mu} \frac{\partial Q_T}{\partial x_{\nu}}
- x_{\nu} \frac{\partial Q_T}{\partial x_{\mu}} \right)
e^{\nu} \wedge e^{n+\mu}, \cr
R_{n+\mu, n+\nu} &= 
- \frac{1}{2(x_{\mu}^2 - x_{\nu}^2)}
\left( x_{\nu} \frac{\partial Q_T}{\partial x_{\mu}}
- x_{\mu} \frac{\partial Q_T}{\partial x_{\nu}}
\right) e^{\mu} \wedge e^{\nu} \cr
& \qq - \frac{1}{2(x_{\mu}^2 - x_{\nu}^2)}
\left( x_{\mu} \frac{\partial Q_T}{\partial x_{\mu}}
- x_{\nu} \frac{\partial Q_T}{\partial x_{\nu}}
\right) e^{n+\mu} \wedge e^{n+\nu}.
\end{split}
\ee
Let $I_{\mu}$ be the differential operator
\begin{equation}
I_{\mu}=\frac{1}{2} \frac{\partial^2}{\partial x_{\mu}^2}+
\sum_{\rho \neq \mu} \frac{1}{x_{\rho}^2-x_{\mu}^2} 
\left( x_{\rho} \frac{\partial}{\partial x_{\rho}}-
x_{\mu} \frac{\partial}{\partial x_{\mu}} \right).
\end{equation}
 The components 
 $\mathcal{R}_{ab}=\sum_{c=1}^{2 n}R^{c}{}_{acb}$ of
 the Ricci curvature are calculated
 as\\
\begin{equation}
\mathcal{R}_{\mu \nu}=\mathcal{R}_{n+\mu, n+\nu}=
-\delta_{\mu \nu}I_{\mu}(Q_{T}).
\end{equation}
Using the expression $Q_T=\sum_{\mu=1}^n (X_\mu/U_\mu)$
we have
\begin{equation}
I_{\mu}(Q_T)=\frac{1}{2}\frac{ X^{''}_{\mu}}{ U_\mu}+
\sum_{\rho \neq \mu}\frac{1}{x_{\rho}^2-x_{\mu}^2}
\left( x_{\rho} \frac{ X^{'}_{\rho}}{U_\rho}+
x_{\mu} \frac{ X^{'}_{\mu}}{U_\mu} \right)
-
\sum_{\rho \neq \mu}\frac{1}{x_{\rho}^2-x_{\mu}^2}
\left( \frac{ X_{\rho}}{U_\rho}+
\frac{ X_\mu}{U_\mu} \right),
\end{equation}
where $X^{'}_{\mu}=dX_\mu/dx_\mu$ and $X^{''}_{\mu}=d^2 X_\mu/dx_\mu^2$.
Thus, the scalar curvature $\mathcal{R}=\sum_{a=1}^{2 n} \mathcal{R}_{aa}$ takes the form
\begin{equation}
\mathcal{R}=- \sum_{\mu=1}^{n}\frac{X_{\mu}^{''}}{U_\mu}.
\end{equation}

%%%%%%%%%%%%%%%%%%%%%%%%%%%%%%%%%%%%%%%
\subsection{Einstein Condition}

We first study a condition $\mathcal{R}$=const.
By (2.10) and the identities
\begin{equation}
\sum_{\mu=1}^n \frac{x_{\mu}^{2(n-1)}}{U_{\mu}} = 1,\qquad \qquad
\sum_{\mu=1}^n \frac{x_{\mu}^{2k}}{U_{\mu}} = 0
\end{equation}
($k=1,2, \cdots, n-2$),~~it is easy to see that the function 
\begin{equation}
X_\mu= \sum_{k=1}^{n} c_{2 k} x_{\mu}^{2 k}+b_{\mu} x_\mu+d_\mu
\end{equation} 
gives a constant scalar curvature $\mathcal{R}=-2n (2 n-1) c_{2n}$
for arbitrary constants $c_{2 k}, b_\mu$ and $d_\mu$.
Conversely, we can show that the condition $\mathcal{R}$=const.
implies (2.12). In fact, from (2.10) one has
\begin{equation}
\mathcal{R} U_\mu =-X_{\mu}^{''}-\sum_{\nu \ne \mu} P^{\nu}_{\mu}
X_{\nu}^{''},
\end{equation}
where
\begin{equation}
P^{\nu}_{\mu}=\frac{U_\mu}{U_\nu}=-
\frac{\prod_{\sigma \ne \mu, \nu} (x_\mu^2-x_\sigma^2)}
{\prod_{\lambda \ne \mu, \nu}(x_\nu^2-x_\lambda^2)}.
\end{equation}
Applying the differential operator $(\partial/\partial x_\mu)^{2 n-1}$
to this relation we obtain
\begin{equation}
\left( \frac{\rmd}{\rmd x_\mu} \right)^{2n+1} X_\mu =0,
\end{equation}
which means that $X_\mu$ must be  polynomials of order $2n$.
Taking $X_\mu$ in the general polynomials of order $2 n$ we infer
from (2.13) that they have the form (2.12). Thus, we have shown
that the scalar curvature is a constant if and only if 
$X_{\mu}$ takes of the form (2.12).

Now, we can examine the Einstein condition, i.e.
$\mathcal{R}_{ab}=\Lambda \delta_{ab}$, where $\Lambda$
represents a cosmological constant. Substituting
(2.12) into (2.9) we obtain
\begin{equation}
\mathcal{R}_{\mu \nu}=\mathcal{R}_{n+\mu, n+\nu}=
\delta_{\mu \nu} \left( -(2n-1)c_{2n}+ K_\mu
\right),
\end{equation}
where
\begin{equation}
K_{\mu}=\sum_{\rho \ne \mu}\frac{1}{x_{\rho}^2-x_{\mu}^2}
\left( \frac{d_{\rho}}{U_\rho}+\frac{d_{\mu}}{U_\mu} \right).
\end{equation}
The Einstein condition requires $K_1=K_2= \cdots =K_n$=const.
This implies $d_1=d_2= \cdots =d_n$, and then $K_\mu=0$. Therefore, denoting the common
value of $d_\mu$ by $c_0$ we reproduce the function $X_\mu$ given in (1.5). 
It should be noticed that
\be
Q_{T} = \sum_{\mu=1}^n
\frac{X_{\mu}}{U_{\mu}}
= c_{2n} \sum_{\mu=1}^n x_{\mu}^2 +c_{2n-2}
+ V,
\ee
where
\be
V= \sum_{\mu=1}^n \frac{b_{\mu} x_{\mu}}{U_{\mu}}.
\ee
 The Ricci curvature is given by
\begin{equation}
\mathcal{R}_{ab} = (2 n-1) \lambda \delta_{ab},
\end{equation}
with $\lambda=-c_{2n}$. From (2.6)
the corresponding curvature two forms are written as $(\mu \neq \nu)$
\be
\begin{split}
R_{\mu\nu}
&= \left[ \lambda - \frac{1}{2(x_{\mu}^2 - x_{\nu}^2)}
\left( x_{\mu} \frac{\partial V}{\partial x_{\mu}}
- x_{\nu} \frac{\partial V}{\partial x_{\nu}} \right)
\right] e^{\mu} \wedge e^{\nu} \cr
& \qq - \frac{1}{2(x_{\mu}^2 - x_{\nu}^2)}
\left( x_{\nu} \frac{\partial V}{\partial x_{\mu}}
- x_{\mu} \frac{\partial V}{\partial x_{\nu}}
\right) e^{n+\mu} \wedge  e^{n+\nu},\cr
R_{\mu, n+\mu} &= 
\left( \lambda - \frac{1}{2} \frac{\partial^2 V}{\partial x_{\mu}^2}
\right) e^{\mu} \wedge e^{n+\mu} \cr
& \qq + \sum_{\rho \neq \mu}
\frac{1}{x_{\mu}^2 - x_{\rho}^2}
\left( x_{\mu} \frac{\partial V}{\partial x_{\rho}}
- x_{\rho} \frac{\partial V}{\partial x_{\mu}} \right)
e^{\rho} \wedge e^{n+\rho},
\qq (\mbox{no sum})\cr
R_{\mu, n+\nu} &= \left[ \lambda
- \frac{1}{2(x_{\mu}^2-x_{\nu}^2)}
\left( x_{\mu} \frac{\partial V}{\partial x_{\mu}}
- x_{\nu} \frac{\partial V}{\partial x_{\nu}} \right)
\right] e^{\mu} \wedge e^{n+\nu} \cr
& + \frac{1}{2(x_{\mu}^2 -x_{\nu}^2)}
\left( x_{\mu} \frac{\partial V}{\partial x_{\nu}}
- x_{\nu} \frac{\partial V}{\partial x_{\mu}} \right)
e^{\nu} \wedge e^{n+\mu}, \cr
R_{n+\mu, n+\nu} &= 
- \frac{1}{2(x_{\mu}^2 - x_{\nu}^2)}
\left( x_{\nu} \frac{\partial V}{\partial x_{\mu}}
- x_{\mu} \frac{\partial V}{\partial x_{\nu}}
\right) e^{\mu} \wedge e^{\nu} \cr
& \qq + \left[ \lambda
- \frac{1}{2(x_{\mu}^2 - x_{\nu}^2)}
\left( x_{\mu} \frac{\partial V}{\partial x_{\mu}}
- x_{\nu} \frac{\partial V}{\partial x_{\nu}}
\right) \right] e^{n+\mu} \wedge e^{n+\nu}.
\end{split}
\ee
If we put $b_\mu=0$ for all $\mu$, then equations represent the
constant curvature space, $R_{ab}= \lambda e^a \wedge e^b $.

%%%%%%%%%%%%%%%%%%%%%%%%%%%%%%%%%%%%%%%%%%%%%%%%%%%%%%%%%%%%%%%%%%%%
\subsection{K\"ahler Condition}

The natural K\"ahler form associated with the metric (1.1) is
\begin{eqnarray}
\omega &=& \sum_{\mu=1}^n e^{\mu} \wedge e^{n+\mu} \nonumber \\
       &=& \sum_{\mu=1}^n \rmd x_\mu \wedge \left( \sum_{k=0}^{n-1} A_\mu^{(k)} \rmd \psi_{k} \right).
\end{eqnarray}
This two-form is not closed, but there exists a scaling limit in which it becomes closed.
Indeed we can take a limit $x_\mu=1+\epsilon \xi_\mu~~( \epsilon \rightarrow 0)$ together with
a suitable transformation of the coordinates $\psi_k$\footnote{
The scaling limit in four-dimension was explicitly given in \cite{MS}.}. Then, we have a closed two-form 
in the form $\omega=\sum_i^{n} \rmd\sigma_i \wedge \rmd t^i$, where $\sigma_i$ are
the elementary symmetric polynomials of $\xi_\mu$'s. Now the metric (1.1) reduces to
the K\"ahler metric presented in \cite{ACG} (see Proposition 11).

%%%%%%%%%%%%%%%%%%%%%%%%%%%%%%%%%%%%%%%%%%%%%%%%%%%%%%%%%%%%%%%%%%%%%%%%%%%%%%%%%%%%%%%%%%%%%%%%%%%%%%%%%%%%%%%
\section{$D=2n+1$}

For the metric (1.2) we introduce the following
orthonormal frame 
$\{ \hat{e}^a \}=\{ \hat{e}^{\mu}, \hat{e}^{n+\mu}, \hat{e}^{2n+1} \}$
$(\mu=1,2,\dotsc, n)$:
\begin{equation}
\hat{e}^{\mu} = e^{\mu},~~
\hat{e}^{n+\mu} = e^{n+\mu},~~
\hat{e}^{2n+1} = \sqrt{S}
\left( \sum_{k=0}^n A^{(k)} \rmd \psi_k \right),
\end{equation}
where $e^{\mu}$ and $e^{n+\mu}$ are defined by (2.1).
The  connection 1-forms $\hat{\omega}_{ab}$ are given by
\begin{eqnarray}
\hat{\omega}_{\mu \nu} &=& \omega_{\mu \nu}, \nonumber \\
\hat{\omega}_{\mu,n+\nu} &=& \omega_{\mu, n+\nu} + \delta_{\mu \nu}
\frac{\sqrt{S}}{x_{\mu}} \hat{e}^{2n+1}, \nonumber \\
\hat{\omega}_{n+\mu, n+\nu} &=& \omega_{n+\mu, n+\nu}, \\
\hat{\omega}_{\mu, 2n+1} &=& \frac{\sqrt{S}}{x_{\mu}} \hat{e}^{n+\mu}
- \frac{\sqrt{Q_{\mu}}}{x_{\mu}} \hat{e}^{2n+1}, \nonumber \\ 
\hat{\omega}_{n+\mu, 2n+1} &=& - \frac{\sqrt{S}}{x_{\mu}} \hat{e}^{\mu} \nonumber,
\end{eqnarray}
with
$\omega_{ab}$ defined by (2.3).
Shifting the arbitrary function $X_\mu$ by 
\begin{equation}
X_\mu=\hat{X}_{\mu}+
\frac{(-1)^{n} c}{x_\mu^2},
\end{equation}
we have
\begin{equation}
Q_T=\hat{Q}_T-S,~~ \hat{Q}_{T}=\sum_{\mu=1}^{n} 
\frac{\hat{X}_{\mu}}{U_{\mu}}.
\end{equation}
Then, the curvature two forms 
$\hat{R}_{\mu \nu}, \hat{R}_{\mu,n+\nu}$ and $\hat{R}_{n+\mu, n+\nu}$
are obtained by
the replacement
$Q_T \rightarrow \hat{Q}_T$ in (2.6),
and
the remaining ones are calculated as
\begin{equation}
\hat{R}_{\mu,2n+1}
=- \frac{1}{2x_{\mu}}
\frac{\partial \hat{Q}_T}{\partial x_{\mu}}
\hat{e}^{\mu} \wedge \hat{e}^{2n+1},~~
\hat{R}_{n+\mu,2n+1}
= - \frac{1}{2x_{\mu}} \frac{\partial \hat{Q}_T}{\partial x_{\mu}}
\hat{e}^{n+\mu} \wedge \hat{e}^{2n+1}.
\end{equation}
The Ricci curvature $\hat{\mathcal{R}}_{ab}$ and the scalar curvature
$\hat{\mathcal{R}}$ are given by 
\begin{equation}
\hat{\mathcal{R}}_{\mu \nu}=\hat{\mathcal{R}}_{n+\mu, n+\nu}
=-\delta_{\mu \nu} \left(
I_{\mu}(\hat{Q}_T) + \frac{1}{2 x_\mu} 
\frac{ \partial \hat{Q}_T}{\partial x_\mu} \right),~~
\hat{\mathcal{R}}_{2 n+1, 2 n+1}=- \sum_{\rho} \frac{1}{ x_{\rho}}
 \frac{\partial \hat{Q}_T}{\partial x_\rho}
\end{equation}
and
\begin{equation}
\hat{\mathcal{R}}=-\sum_{\mu=1}^n \frac{\hat{X}_{\mu}^{''}}{U_\mu}
-2 \sum_{\mu=1}^n \frac{1}{x_\mu}\frac{\hat{X}_{\mu}^{'}}{U_\mu}.
\end{equation}

%%%%%%%%%%%%%%%%%%%%%%%%%%%%%%%%%%%%
\subsection{Einstein Condition}

Using similar arguments to the case of $D=2n$, we can show that
the scalar curvature is a constant if and only if $X_{\mu}$
takes of the form
\begin{equation}
X_{\mu}=\sum_{k=1}^{n} c_{2k} x_{\mu}^{2k}+b_\mu + \frac{d_\mu}{x_\mu}
+\frac{(-1)^{n}c}{x_\mu^2}.
\end{equation}
Then the components of the Ricci curvature are
\begin{eqnarray}
\hat{\mathcal{R}}_{\mu \nu}= \hat{ \mathcal{R}}_{n+\mu, n+\nu}&=&
-\delta_{\mu \nu} \left(
2 n c_{2n}+ \frac{1}{2}\frac{d_\mu}{x_\mu^3 U_\mu}-\sum_{\rho \ne \mu}
\left( \frac{d_\rho}{x_\rho U_\rho}+\frac{d_\mu}{x_\mu U_\mu} \right) \right), \nonumber \\
\hat{\mathcal{R}}_{2n+1, 2n+1}&=&-2 n c_{2n}+\sum_{\mu=1}^{n}\frac{d_\mu}{x_\mu^3 U_\mu},
\end{eqnarray}
which satisfy the Einstein condition if and only if $d_\mu$ vanishes for all $\mu$.
Thus we reproduce the function $X_\mu$ given in (1.6).
Now $\hat{\mathcal{R}}_{ab}= 2 n \lambda \delta_{ab}$ with $\lambda=-c_{2n}$ and
$Q_T$ is given by
\be
Q_T = c_{2n} \left( \sum_{\mu=1}^n x_{\mu}^2 \right) +c_{2n-2}+\hat{V}- S,
\ee
where
\be
\hat{V} = \sum_{\mu=1}^n \frac{b_{\mu}}{U_{\mu}}.
\ee
By substituting these expressions into (2.6) and (3.5), we  obtain (2.21) replaced
$V$ with $\hat{V}$, and
\begin{equation}
\hat{R}_{\mu, 2n+1}
= \left( \lambda - \frac{1}{2x_{\mu}} \frac{\partial \hat{V}}{\partial x_{\mu}}
\right) e^{\mu} \wedge \hat{e}^{2n+1},~~
\hat{R}_{n+\mu, 2n+1}
= \left( \lambda - \frac{1}{2x_{\mu}} \frac{\partial \hat{V}}{\partial x_{\mu}}
\right) e^{n+ \mu} \wedge \hat{e}^{2n+1}.
\end{equation}
If we choose all of $b_\mu$ as an equal value, then $\hat{V}=0$ and hence 
the equations represent
the constant curvature space.

%%%%%%%%%%%%%%%%%%%%%%%%%%%%%%%%%%%%%%%%%%%%%%%%%%%%%%%%%%%%%%%%%%%%%%%%%%%%%%%%%%%%%%%%%%%%%%%%%%%%%%%%%%%
\section{Concluding Remarks}

We have explicitly calculated the Riemannian curvature corresponding to
the metrics (1.1) and (1.2). The components have a compact expression
by introducing the single function $Q_T$.  We have also proved that the conditions (1.5) and (1.6)
lead to the Einstein metrics for all dimensions.

Finally we comment on type D condition \cite{CMPP}.
For fixed $\mu$ let us define the complex vector fields\footnote{The vector fields $e_{a}$ are
dual to the 1-forms $e^a$ given in (2.1) and (3.1).}
\begin{equation}
k=Q_\mu^{-1/2}(e_\mu+ \rmi e_{n+\mu})/\sqrt{2},~~
\ell=Q_\mu^{1/2}(e_\mu- \rmi e_{n+\mu})/\sqrt{2}.
\end{equation}
Then $\{k, \ell, e_{\alpha} \}(\alpha \ne \mu, n+\mu)$ gives a null orthonormal frame 
for (1.1) or (1.2) :
\begin{equation}
\langle k,k \rangle=\langle \ell,\ell \rangle=0,~~
\langle k,\ell \rangle =1,~~
\langle k,e_\alpha \rangle=\langle \ell,e_\alpha \rangle=0,~~
\langle e_\alpha, e_\beta\rangle=\delta_{\alpha,\beta}.
\end{equation}
The covariant derivatives are given by
$\nabla_{e_b}\,e_a=\omega_{c a}(e_b) e_c$,~which can be easily calculated by
(2.3) and (3.2). Especially, we have
\begin{equation}
\nabla_{k}\,k=0,
\end{equation}
which means that the integral curve of $k$ is a geodesic. It is easy
to confirm that the Weyl curvature satisfies the type D condition: 
\begin{eqnarray}
W(k, e_{\alpha}, e_{\beta}, e_{\gamma})&=& W(\ell,e_{\alpha},e_{\beta},e_{\gamma})=0,~~
W(k, e_{\alpha}, k, e_{\beta})=W(\ell, e_{\alpha}, \ell, e_{\beta})=0,\nonumber\\
W(k,\ell,k,e_{\alpha})&=& W(k,\ell,\ell,e_{\alpha})=0.
\end{eqnarray}

%%%%%%%%%%%%%%%%%%%%%%%%%%%%%%%%%%%%%%%%
{\bf{Acknowledgements}}

This work is supported by the 21 COE program
``Construction of wide-angle mathematical basis focused on knots".
The work of Y.Y is supported by the Grant-in Aid for Scientific
Research (No. 17540262 and No. 17540091)
from Japan Ministry of Education. 
The work of T.O is supported by the Grant-in Aid for Scientific
Research (No. 18540285)
from Japan Ministry of Education.

%%%%%%%%%%%%%%%%%%%%%%%%%%%%%%%%%%%%%%%%%%%%%%%%%%%%%%%%%%%%%%%%%%%%%%%%%%%%%%%%%%%%%%%%%%%%%%%%%%%%%%%%%%%%%%%%%%%5


\begin{thebibliography}{99}

\bibitem{MP}
R.C. Myers and M.J. Perry,
``Black holes in higher dimensional space-times,"
Ann. Phys. {\bf{172}}, 304 (1986).

\bibitem{HHT}
S.W. Hawking, C.J. Hunter and M.M. Taylor-Robinson,
``Rotation and the AdS/CFT correspondence,"
Phys. Rev. {\bf{D59}}, 064005 (1999),
\texttt{hep-th/9811056}.

\bibitem{GLPP1}
G.W. Gibbons, H. L\"{u}, D.N. Page and C.N. Pope,
``The general Kerr-de Sitter metrics in all dimensions,"
J. Geom. Phys. {\bf{53}}, 49 (2005),
\texttt{hep-th/0404008}.

\bibitem{GLPP2}
G.W. Gibbons, H. L\"{u}, D.N. Page and C.N. Pope,
``Rotating black holes in higher dimensions with a cosmological constant,"
Phys. Rev. Lett. {\bf{93}}, 171102 (2004),
\texttt{hep-th/0409155}.

\bibitem{CLP1}
W. Chen, H. L\"{u} and C.N. Pope,
``Kerr-de Sitter Black Holes with NUT Charges,"
\texttt{hep-th/0601002}.

\bibitem{CLP2}
W. Chen, H. L\"{u} and C.N. Pope,
``General Kerr-NUT-AdS Metrics in All Dimensions,"
Class. Quant. Grav. {\bf{23}}, 5323 (2006),
\texttt{hep-th/0604125}.

\bibitem{HSY}
Y. Hashimoto, M. Sakaguchi and Y. Yasui,
``Sasaki-Einstein Twist of Kerr-AdS Black Holes,"
Phys. Lett. {\bf{B 600}}, 270 (2004),
\texttt{hep-th/0407114}.

\bibitem{CLPP1}
M. Cveti\v{c}, H. L\"{u}, D.N. Page and C.N. Pope,
``New Einstein-Sasaki spaces in five and higher dimensions,"
Phys. Rev. Lett. {\bf{95}}, 071101 (2005),
\texttt{hep-th/0504225}.

\bibitem{CLPP2}
M. Cveti\v{c}, H. L\"{u}, D.N. Page and C.N. Pope,
``New Einstein-Sasaki and Einstein spaces from Kerr-de Sitter,"
\texttt{hep-th/0505223}.

\bibitem{OY}
T. Oota and Y. Yasui,
``Explicit Toric Metric on Resolved Calabi-Yau Cone,"
Phys. Lett. {\bf{B 639}}, 54 (2006),
\texttt{hep-th/0605129}.

\bibitem{LP}
H. L\"{u} and C.N. Pope,
``Resolutions of Cones over Einstein-Sasaki Spaces," \\
\texttt{hep-th/0605222}.

\bibitem{MS}
D. Martelli and J. Sparks,
``Toric Sasaki-Einstein metrics on $S^2 \times S^3$,"
Phys. Lett. \textbf{B621}, 208 (2005),
\texttt{hep-th/0505027}.

\bibitem{ACG}
V. Apostolov, D.M.J. Calderbank and P. Gauduchon,
``Hamiltonian $2$-Forms in K\"{a}hler Geometry, I General Theory,"
J. Diff. Geom. \textbf{73}, 359 (2006), \\
\texttt{math.DG/0202280}.

\bibitem{CMPP}
A. Coley, R. Milson, V. Pravda and A. Pravdova,
``Classification of the Weyl Tensor in Higher Dimensions,"
Class. Quant. Grav. {\bf{21}}, L35 (2004),
\texttt{gr-qc/0401008}.



\end{thebibliography}
\end{document}